\documentclass[12pt]{article}
\usepackage{pictex} \usepackage{graphics} \usepackage{amsmath} 
\usepackage{amssymb} \usepackage{amsthm} %\usepackage{materi} 
\newcommand{\nc}{\newcommand}  
\nc{\di}{\displaystyle} \nc{\nn}{\nonumber} 
\newcounter{zahl}
\newenvironment{zael}{\begin{list}
{{\rm \arabic{zahl})}\hfill}{\usecounter{zahl}
\labelwidth0.6cm \leftmargin0.8cm \labelsep0.2cm
\rightmargin0cm \parsep1ex plus0.2ex minus0.1ex \itemsep0ex
plus0.2ex}}{\end{list}}
\nc{\ns}{\normalsize} \nc{\seq}{\subseteq} 
 \renewcommand{\Im}{\mathrm{Im}\,} 
\nc{\AH}{\mathcal{A}} \nc{\BE}{\mathcal{B}} \nc{\CE}{\mathcal{C}} 
\nc{\D}{\mathcal{D}} \nc{\E}{\mathcal{E}} \nc{\EF}{\mathcal{F}} 
\nc{\GE}{\mathcal{G}} \nc{\HA}{\mathcal{H}} \nc{\J}{\mathcal{J}} 
\nc{\KA}{\mathcal{K}} \nc{\EL}{\mathcal{L}} \nc{\PE}{\mathcal{P}} 
\nc{\ER}{\mathcal{R}} \nc{\ES}{\mathcal{S}} \nc{\TE}{\mathcal{T}} 
\nc{\EM}{\mathcal{M}} \nc{\EN}{\mathcal{N}} \nc{\OH}{\mathcal{O}} 
\nc{\U}{\mathcal{U}} \nc{\WE}{\mathcal{W}} \nc{\EX}{\mathcal{X}} 
\nc{\Y}{\mathcal{Y}} %\nc{\V}{\mathcal{V}} \nc{\ZE}{\mathcal{Z}} 
\nc{\ma}[1]{\mbox{$\,{#1}\,$}} \nc{\ek}{\protect\\[1ex]} 
\nc{\zx}{\protect\\[2ex]}

 \newcommand{\R}{{\mathbb R}} 
  
 \nc{\IR}{\mbox{\bf R}} \nc{\IN}{\mbox{\bf N}} 
\nc{\ZZ}{\mbox{\bf Z}} \nc{\la}{\lambda} \nc{\La}{\Lambda} 
\nc{\da}{\delta} \nc{\Da}{\Delta} \nc{\ta}{\theta} \nc{\Ta}{\Theta} 
\nc{\na}{\nabla} \nc{\ue}{\infty} \nc{\vp}{\varphi} \nc{\vta}{\vartheta} 
\nc{\Gm}{\Gamma} \nc{\gm}{\gamma} \nc{\ka}{\kappa} \nc{\si}{\sigma} 
\nc{\Si}{\Sigma} \nc{\al}{\alpha} \nc{\be}{\beta} \nc{\om}{\omega} 
\nc{\Om}{\Omega} \nc{\pa}{\partial} \nc{\ti}{\times} \nc{\n}{|} 
\nc{\rub}{\,\rule[-2.7pt]{.02in}{4mm}\,} \nc{\ab}{\|} \nc{\s}{\tilde} 
\nc{\ve}{\varepsilon} \nc{\fa}{\forall} \nc{\ov}{\overline} 
\nc{\un}{\underline} \nc{\Llr}{\Longleftrightarrow} 
\nc{\llr}{\longleftrightarrow} \nc{\ra}{\rightarrow} 
\nc{\lra}{\longrightarrow} \nc{\rh}{\rightharpoonup} \nc{\Ra}{\Rightarrow} 
\nc{\ran}{\rangle} \nc{\lan}{\langle} \nc{\bs}{\backslash} \nc{\ko}{\,,\,} 
\nc{\eq}[1]{\mbox{\rm {(\ref{E#1})}}} 
\nc{\ha}{\frac{1}{2}} 
\nc{\kla}{\,[\,} \nc{\klz}{\,]\,} \nc{\lk}{\left[} \nc{\rk}{\right]} 
\nc{\lb}{\left\{} \nc{\rb}{\right\}} \nc{\rr}{\right)} \nc{\lr}{\left(} 
\nc{\f}{\big(} \nc{\g}{\big)} \nc{\Ba}{\Big(} \nc{\Bz}{\Big)} 
\nc{\Bka}{\Big[} \nc{\Bkz}{\Big]} \nc{\bka}{\big[} \nc{\bkz}{\big]} 
\nc{\Blb}{\Big\{} \nc{\Brb}{\Big\}} \nc{\blb}{\big\{} \nc{\brb}{\big\}} 
\nc{\pn}{\par\noindent} \nc{\emp}{\emptyset} \nc{\Ri}{\Rightarrow} 
\nc{\hph}{\hphantom} \nc{\vph}{\vphantom} 
\nc{\vpn}{\vspace{2ex}\par\noindent} \nc{\vpar}{\vspace{2ex}\par} 
\nc{\mathe}[1]{\mbox{${\di {#1}}$}} \nc{\equno}[1]{\\[-.5ex] 
\mbox{}\label{#1}\\[-.5ex]} \nc{\tr}{{\mathrm{tr}}\,} 
\newtheorem{lem}{Lemma} 
\newtheorem{theo}[lem]{Theorem}

\begin{document} 
\title{{\Large {\bf Wave scattering by many small particles embedded in a 
medium. }}} \author{A. 
G. Ramm \\ {\ns (Mathematics Department, Kansas State University,}\\ {\ns 
Manhattan, KS66506, USA} \\ {\ns and TU Darmstadt, Germany)}\\ {\small 
ramm@math.ksu.edu}} \date{} 
\maketitle 
\begin{abstract}\noindent 
Theory of scattering by many small bodies is developed under various
assumptions concerning the ratio $\frac{a}{d}$, where $a$ is the
characteristic dimension of a small body and $d$ is the distance between
neighboring bodies $d = O(a^{\ka_1})$, $0 < \ka_1 < 1$. On the boundary
$S_m$ of every small body an impedance-type condition is assumed $u_N =
\zeta_m u$ on $S_m$, $1 \leq m \leq M$, $\zeta_m = h_m a^{-\ka}$, $0 <
\ka$, $h_m$ are constants independent of $a$. The behavior of the field
in the region in which $M = M(a)\gg 1$ small particles are embedded is
studied as $a\ra 0$ and $m(a)\ra \ue$. Formulas for the refraction
coefficient of the limiting medium are derived 
 under the assumptions:
a) $ \ka_1=(2-\ka)/3$, $0<\ka\leq 1$, and b) $\ka_1=1/3$,  $\ka>1$.
\end{abstract}
\pn
{\small {\sc PACS}: 43.20. + g, 62.40. + d: 78.20. - e.  } 

\section{Introduction}\label{S1}
The theory of wave scattering by small
bodies was originated by Rayleigh in 1871 \cite{L}. In \cite{R476} this
theory was developed for small bodies of arbitrary shapes, analytic
formulas for the $S$-matrix for acoustic and electromagnetic (EM) wave
scattering by small bodies of arbitrary shapes have been derived. These
formulas allow one to calculate the $S$-matrix with any desired
accuracy. Analytic formulas for the electric and magnetic polarizability
sensors have been derived for bodies of arbitrary shapes \cite{R117},
\cite{R476}. In \cite{R450} -- \cite{R533} a theory of wave scattering by
many small bodies embedded in a bounded domain filled in by a material
with known properties was developed. It was assumed in \cite{R528} and
\cite{R529} that the characteristic size of the small particles (bodies)
is $a$, that the distance $d$ between two neighboring particles is
is of the order $d = O(a^{1/3})$, that
the total number of the embedded particles $M = O(\frac{1}{a})$, and 
that the
boundary condition on the boundary $S_m$ of $m$-th particle $D_m$ is of
impedance type:
\begin{equation}
\label{E1}
u_N = \zeta_M u\quad \mbox{on }S_m,\quad 1 \leq m \leq M,
\end{equation}
where $N$ is the unit normal to $S_m$ directed out of $D_m$, and
$\zeta_m = \frac{h_m}{a}\,,$ where $h_m$, $\Im\, h_m \le 0$, is a constant
independent of $a$.

The waves in the original material are described by the equation
\begin{equation}
\label{E2}
L_0 u_0:= \bka \na^2 + k^2 n_0^2(x)\bkz u_0 = 0\quad \mbox{in }\R^3,
\end{equation}
where
\begin{equation}
\label{E3}
n^2_0(x) = 1\quad \mbox{in }D' = \R^3\bs D,
\end{equation}
$D$ is a bounded domain, and $n_0^2(x)$ is continuous in $D$ (or
piecewise-continuous with a finite number of discontinuities, which are
smooth surfaces), $\Im n_0^2 \geq 0$. The scattering solution to \eq{2}
satisfies the radiation condition
\begin{eqnarray}
\label{E4}
& u_0 = e^{ik\al\cdot x} + v_0, &\\
&\displaystyle \frac{\pa v_0}{\pa r}-ik v_0 = o\f \frac{1}{r}\g,\quad r:=
\n x\n\ra \ue.&\label{E5}
\end{eqnarray}
If small particles are embedded in $D$, then the scattering problem
consists of finding the solution to the following problem:
\begin{eqnarray}
\label{E6} L_0 u_M &= & 0\quad \mbox{in }\R^3\bs \bigcup^M_{m=1} D_m, \\
\frac{\pa u_M}{\pa N} & = & \zeta_m u_M\quad \mbox{on } S_m,\quad 1 \leq
m \leq M,\label{E7}\\
u_M & = & u_0 + v_M,\label{E8}
\end{eqnarray}
where $u_0$ solves problem \eq{2}, \eq{4}, \eq{5} and $v_M$ satisfies
the radiation condition similar to \eq{5}.
\par
It is proved in \cite{R529} that problem \eq{6} -- \eq{8} has a unique
solution and this solution is of the form
\begin{equation}
\label{E9}
u_M = u_0(x) + \sum^M_{m=1}\int_{S_m}G(x,t)\, \si_m(t)dt,
\end{equation}
where $G(x,y)$ is the Green function of the operator $L_0$ 
for $M=0$, i.e., in the absence
of small particles:
\begin{equation}
\label{E10}
L_0 G(x,y) = -\da(x-y)\quad \mbox{in }\R^3,
\end{equation}
$G$ satisfies the radiation condition \eq{5}, and $\si_m$ solves the
equation
\begin{equation}
\label{E11}
u_{e_N}-\zeta_m u_e + \frac{A_m\si_m-\si_m}{2} - \zeta_m T_m\si_m =
0\quad \mbox{on }S_m.
\end{equation}
Here $u_e$ is the effective field acting on the $m$-th particle:
\begin{equation}
\label{E12}
u_e(x):= u_e^{(m)}(x):= u_M(x) - \int_{S_m}G(x,t)\, \si(t)dt,\quad x\in \R^3,
\end{equation}
\begin{equation}
\label{E13}
A_m\si_m:= 2 \int_{S_m} \frac{\pa G(s,t)}{\pa N_s}\, \si_m(t)dt,\quad
T_m\si_m:= \int_{S_m}G(s,t)\si_m(t)dt.
\end{equation}
It was proved in \cite{R529} that
\begin{equation}
\label{E14}
G(x,y) = \frac{1}{4\pi\n x-y\n}\, \bka 1 + O(\n x-y\n )\bkz,\quad \n
x-y\n \ra 0,
\end{equation}
and one can differentiate formula \eq{14}. 

The following result is also
proved in \cite{R529}. Assume that $D_m$ is a ball of radius $a$
centered at a point $x_m$. Let $h(x)$ be an arbitrary continuous
function in $D$, $\Im h(x)\leq 0$,  $\Da_p\seq D$ be any subdomain of
$D$, and $\EN(\Da_p)$ be the number of particles in $\Da_p$. Assume that

\begin{equation}
\label{E15}
\EN(\Da_p) = \frac{1}{a} \int_{\Da_p} N(x)dx \bka 1 + o(1)\bkz,\quad a
\ra 0,
\end{equation}
where $N(x)\geq 0$ is a given continuous function in $D$. Let
\begin{equation}
\label{E16}
p(x):= \frac{4\pi \, N(x)\, h(x)}{1 + h(x)}\,.
\end{equation}
Finally, assume that $\zeta_m:= \frac{h(x_m)}{a}\,$.
Now the result can be formulated:
\begin{theo}[\cite{R529}]\label{T1}
Under the above assumptions there exists the limit
\begin{equation}
\label{E17}
\lim_{a\ra 0}\ab u_e(x) - u(x)\ab_{C(D)} = 0.
\end{equation}
The function $u(x)$ solves the problem
\begin{align}
\label{E18}
Lu & := [\na ^2 + k^2 - q(x)] u = 0 \quad \mbox{in }\R^3,\\
\label{E19} u& = u_0 + v,
\end{align}
where $u_0$ satisfies equations \eq{2}, \eq{4}, \eq{5}, 
the function $v$ satisfies the 
radiation
condition siilar to \eq{5}, and
\begin{equation}
\label{E20}
q(x):= q_0(x) + p(x),\quad n^2(x):= 1-k^{-2} q(x),
\end{equation}
where $p(x)$ is defined in \eq{16},
\begin{equation}
\label{E21}
q_0(x):= k^2 - k^2 n_0^2(x),
\end{equation}
and $n_0^2(x)$ is the coefficient in \eq{2}.
\end{theo}
{\it The aim of this paper is to investigate the behavior of $u_e(x)$ when
the assumptions $\zeta_m = \frac{h(x_m)}{a}\,$, $d=O\f a^{1/3}\g$, $M =
O\f \frac{1}{a}\g$ are replaced by the following more general assumptions:
\begin{equation}
\label{E22}
\zeta_m = \frac{h(x_m)}{a^\ka}\,,\quad d = O\f a^{\ka_1}\g,\quad M = O\f
\frac{1}{a^{3\ka_1}}\g, 
\end{equation}
where $\ka> -1 $ and $0\leq \ka_1 <1$ are parameters.}

 If $\ka_1 = 1$,
then the distance between neighboring particles is of the order of the
size of a small particle. This is a special case which is not covered
by a rigorous theory. However, if $\ka_1$ is close to $1$, then 
practically the distance between neighboring particles is
very close to the order $a$ of the size of a small particle.

In \cite{R529} the theory was developed in detail in the case $\ka = 1$, 
$\ka_1 =\frac{1}{3}$. 

The questions we are interested in this
paper are:
\begin{zael}
\item For what ranges of $\ka$ and $\ka_1$ the limit $u(x) $ of $u_e(x)$, 
as $a\ra 0$, does exist?
\item What is the equation which this limit $u(x)$ solves?
\end{zael}
The answers we give are:
\begin{zael}
\item If $\ka < 1$ and $\ka_1 = \frac{2-\ka}{3}\,$, then the limit
\eq{17} exists,
$$\si_m = - \frac{h(x_m)u_e(x_m)}{a^\ka}\, \f 1 + o(1)\g,\quad Q_m = -
4\pi h(x_m)a^{2-\ka}
u_e(x_m)$$
and the limiting function $u(x)$ solves the following equation:
\begin{equation}
\label{E23}
u(x) = u_0(x) -4\pi \int_D G(x,y) h(y)N(y) u(y) dy.
\end{equation}
Therefore, $u$ solves equation \eq{18} with $q(x)$ given by \eq{20},
$q_0(x)$ given by \eq{21}, and
\begin{equation}
\label{E24}
p(x) = 4\pi h(x)N(x),
\end{equation}
where $N(x)\geq 0$ is defined by the formula:
\begin{equation}
\label{E25}
\EN(\Da_p) = \frac{1}{a^{3\ka_1}}\int_{\Da_p}N(x)dx [1 + o(1)],\quad a
\ra 0.
\end{equation}
\item % (2)
If $\ka > 1$ then, as $a\ra 0$, 
$$Q_m = -4\pi\, u_e(x_m) a\f 1 + o(1)\g, \quad
\si_m = - \frac{u_e(x_m)}{a}\, \f 1 + o(1)\g,$$ 
and the limit \eq{17} exists
if $\ka_1 = \frac{1}{3}\,$. The limiting function $u(x)$ solves 
equation \eq{23}
with $h(y) = 1$ and $N(x)$ defined by \eq{25}. The function $u(x)$ also
solves equation \eq{18} with $q(x)$ given by \eq{20}, $q_0(x)$ given by
\eq{21}, and $p(x)$ given by \eq{24} with $h(x) = 1$.
\end{zael}
In both cases, $0<\ka<1$ and  $\ka>1$, we have $\ka_1 <2/3$. This implies
that the total volume of the embedded particles tends to zero as $a\to 0$.
Indeed, the order of the total number of the embedded particles is 
$O(a^{-3\ka_1 })$, and the total volume of the embedded particles is 
of the order $O(a^{3-3\ka_1})\to 0$ as $a\to 0$. 
$$ $$

Let us make a remark about the case when $\ka_1 = \frac{2-\ka}{3}=1\,$. 
In this case $\ka = -1$ and $\zeta_m = h(x_m)a$. Moreover, one has:
$$Q_m \sim a^3\bka \frac{4\pi}{3}\,\Delta u_e(x_m)-4\pi h(x_m)\,
u_e(x_m)\bkz,$$
where $\Delta=\nabla^2$ is the Laplacean, and
$$\si_m \sim u_{e_N} - hau_e(x_m).$$
The quantity $I_m:=|G(x,y_m)Q_m| = O\f a^{3-\ka_1}\g,$
as $ a \ra 0$, and
$$J_m =|\int_{S_m}\bka G(x,t)-G(x,x_m)\bkz \, \si_m(t)dt|= O\big( 
a^{2-2\ka_1} a^2\big),\quad a
\ra 0.$$
For  the relation $J_m\ll I_m$ to hold as $a\to 0$, it is sufficient that
the relation
$$ a^{4-2\ka_1}\ll a^{3-\ka_1}$$
holds. For this relation to hold it is sufficient to have $\ka_1 < 1$.

The relation $J_m\ll I_m$ allows us to use formula (35),  see below,
i.e., approximate the exact formula (31) by an approximate
formula (35) with an  
error which tends to zero as $a\to 0$.
\par
Assuming $\ka_1 < 1$, one has
\begin{equation}
\label{E26}
u_e(x) = u_0 + \sum^M_{m=1}G(x,y^{(p)})\bka \frac{4\pi}{3}\, \Da u_e
(y^{(p)})-4\pi h(y^{(p)})\, u_e(y^{(p)})\bkz a^3\EN(\Da_p).
\end{equation}
We have:
\begin{equation}
\label{E27}
a^3 \EN(\Da_p)= \frac{a^3}{a^{3\ka_1}}\int_{\Da_p}N(x)dx [1+o(1)] \approx
a^{3-3\ka_1}\, N(y^{(p)})\n \Da_p\n,
\end{equation}
where $o(1)$ tends to zero as $a\to 0$.
For the limit of the sum in \eq{26} to exist as $a\ra 0$, it is
necessary and sufficient that $3=3\ka_1$, i.e., $\ka_1 = 1$.
If $\ka_1 = 1$, then the limit 
of $u_e(x)$, as
$a\ra 0$ and $\max_p\, {\rm diam}\, \Da_p \ra 0$, is the function
$u(x)$, which solves the equation
\begin{equation}
\label{E28}
u(x) = u_0(x) + \int_D G(x,y)\, \bka \frac{4\pi}{3}\, \Da u(y)-4\pi
h(y)\, u(y)\bkz\, N(y)dy.
\end{equation}
Applying operator $L_0 = \na^2 + k^2 - q_0(x)$ to \eq{28} and using
equation \eq{10}, one gets
\begin{equation}
\label{E29}
L_0 u = - \bka \frac{4\pi}{3} \, \Da u - 4\pi h(x)\, u(x)\bkz N(x).
\end{equation}
Thus
\begin{equation}
\label{E30}
\bka 1 + \frac{4\pi}{3}\, N(x) \bkz \, \na^2 u + k^2 u-q_0(x) u- 4\pi
h(x)N(x)u(x) = 0.
\end{equation}
The solution $u9x)$ to equations (18) or (30) is a locally $H^2_{loc}$
 function, where $H^2_{loc}$ is the Sobolev space of 
twice differentiable in $L^2-$sense functions on every bounded open subset 
of $\R^3$. This local smoothenss:   $u\in 
H^2_{loc}(\R^3)$,  follows from known results on elliptic regularity,
provided that the coefficients $q_0(x)$ and $h(x)N(x)$ are in $L^2_{loc}$.
If these coefficients are smoother, then $u$ is smoother.

%If $\ka_1=1$ then the distance between the small particle is of the
%order of the size of these particles.\vpar
The assumption  $\ka_1<1$ allows us to prove that formula (35) 
of Section 2 is a good approximation of $u$ as $a\to 0$.

The conclusions, obtained under the assumption  $\ka_1=1$
are not proven to be exact in the limit $a\to 0$.

In Section \ref{S2} we prove the results listed in the answers.
\section{Proofs}\label{S2}
In the proofs we use some  arguments from \cite{R529}.\ek
Case 1). \
Consider first the case $\ka < 1$. Let us write the exact formula
\eq{9} as follows:
\begin{equation}
\label{E31}
u_M(x) = u_0(x) + \sum^M_{m=1}G(x,x_m) Q_m + \sum^M_{m=1} \int_{S_m} G(x,t)\si_m(t)dt,
\end{equation}
where $x_m$ is the center of the ball $D_m$ and
\begin{equation}
\label{E32}
Q_m:= \int_{S_m}\si_m(t)dt.
\end{equation}
One has the following estimates (see \cite{R529}):
\begin{equation}
\label{E33}
\n G(x,y)\n \leq \frac{c}{\n x-y\n}\,,\quad \n \na G(x,y)\n \leq c\,
\max\Ba \frac{k}{\n x-y\n}\,,\: \frac{1}{(x-y)^2}\Bz,
\end{equation}
where $c > 0$ stands for various constants independent of $a$. 

Let us
estimate $Q_m$. Eventually we want to derive sufficient condition for
the relation
\begin{equation}
\label{E34}
I_m:=\n G(x,x_m)Q_m\n \gg \Bigl\n\int_{S_m}\bka G(x,t)-G(x,x_m)\bkz
\si_m(t)dt\Bigr\n := J_m
\end{equation}
to hold as $a\ra 0$ and $\n x-x_m\n \gg a$. This relation allows one to
rewrite the exact formula \eq{31} as an approximate formula:
\begin{equation}
\label{E35}
u_M = u_0(x) + \sum^M_{m=0} G(x,y_m)Q_m,\quad \n x-x_m\n \gg a,
\end{equation}
the error of which tends to zero as $a\to 0$.

To derive a formula for  $Q_m$, integrate \eq{11} over $S_m$ and use the 
divergence theorem to
get:
\begin{equation}
\label{E36}
\frac{4}{3}\, \pi a^3 \, \Da u_e(x_m)- \frac{h(x_m)}{a^\ka}\, u_e(x_m)
4\pi a^2 = Q_m + \frac{h}{a^\ka} \int_{S_m}dx\, \int_{S_m}\,
\frac{\si_m(t)}{4\pi \n s-t\n}.
\end{equation}
One has
\begin{equation}
\label{E37}
\int_{S_m}ds \int_{S_m}\frac{\si_m(t)at}{4\pi \n s-t\n} = \int_{S_m}
dt\, \si_m(t) \int_{S_m}\frac{ds}{4\pi \n s-t\n} = a Q_m.
\end{equation}
Here we have used the formula
\begin{equation}
\label{E38}
\int_{S_m:= \{ s: \n s-x_m\n = a\}} \frac{ds}{4\pi |s-t|}=a,\quad t\in 
S_m.
\end{equation}
Thus, \eq{36} yields:
\begin{equation}
\label{E39}
Q_m = \frac{\frac{4}{3}\, \pi a^3 \, \Da u_e (x_m) - 4\pi h(x_m)
u_e(x_m)a^{2-\ka}}{1 + ha^{1-\ka}}\,.
\end{equation}
If $\ka < 1$ and $a \ra 0$, then \eq{39} implies 
\begin{equation}
\label{E40}
Q_m= - 4\pi h(x_m) u_e(x_m)a^{2-\ka}[1+o(1)],\quad a \ra 0.
\end{equation}
This is the formula for $Q_m$ which we wanted to derive.

If $a\ll 1$, then a formula for $\si_m$ can be derived as follows. The 
function
$u_e$ does not change  at a small distance of order $a$. Therefore one can
assume that in a neighborhood of $D_m$ the function $u_e$ is a constant, 
and one considers a static problem of finding $\si_m$:
\begin{equation}
\label{E41}
u_M(x)= u_e(x_m) + \int_{S_m}\frac{\si_m(t)dt}{4\pi |x-t|}\,,\quad
\frac{\pa u_M}{\pa N} = \zeta_m u_M \mbox{ on }S_m.
\end{equation}
We look for the solution $\si_m = ca^\gm$, where $c$ and $\gm$ are
constants. In this case we have:
\begin{equation}
\label{E42}
\int_{S_m} \frac{\si_m(t)dt}{4\pi\n s-t\n} = ca^\gm \frac{a^2}{\n
x-x_m\n}\,,\quad \n x-x_m\n =O(a ).
\end{equation}
Using the impedance boundary condition \eq{41} and choosing the origin at 
the
point $x_m$, one gets
$$-ca^\gm\, \frac{a^2}{a^2} = \frac{h(x_m)}{a^\ka} \bka u_e(x_m) +
ca^\gm\, \frac{a^3}{a}\bkz.$$
From the above equation one derives:
\begin{equation}
\label{E43}
ca^\gm = - \frac{h(x_m)\, u_e(x_m)}{a^\ka [ a + h(x_m)a^{1-\ka}]}\,.
\end{equation}
If $\ka < 1$ and $a \ra 0$, then equation \eq{43} implies $\gm = -\ka$ and 
$c =
-h(x_m) u_e(x_m)$, so
\begin{equation}
\label{E44}
\si_m = - \frac{h(x_m)u_e(x_m)}{a^\ka}\, [1 + o(1)],\quad a\ra 0.
\end{equation}
This is the formula for $\si_m$ whih we wanted to derive.

Let us find sufficient conditions for the relation \eq{34} to hold. Using
estimates \eq{33} and \eq{39}, we get
\begin{equation}
\label{E45}
G(x,x_m)Q_m = O\Ba \frac{a^{2-\ka}}{a^{\ka_1}}\Bz,\quad a \ra 0,\: \n
x-x_m\n \geq d = O(a^{\ka_1}).
\end{equation}
Using \eq{33} and \eq{44} one gets:
\begin{equation}
\label{E46}
J_m = O\Ba \frac{a}{a^{2\ka_1}}\, a^{2-\ka}\Bz = O(a^{3-\ka -
2\ka_1}),\quad a \ra 0,\: \n x-x_m \n \geq d.
\end{equation}
For \eq{34} to hold it is sufficient to have:
\begin{equation}
\label{E47}
a^{3-\ka - 2\ka_1}\ll a^{2-\ka-\ka_1},\quad a \ra 0.
\end{equation}
{\it This relation holds if $\ka_1 < 1$.} 

Thus, let us assume that $\ka < 1$
and $\ka_1 < 1$, and use formulas \eq{36}, \eq{39} and \eq{40} to get
\begin{equation}
\label{E48}
u_M(x) = u_0(x) - \sum^M_{m=1}G(x,x_m)\, 4\pi h(x_m) u_e(x_m)
a^{2-\ka},\quad \n x-x_m\n \geq d.
\end{equation}
Now we want to pass to the limit $a\to 0$ in equation \eq{48}. To do this,
let us partition the domain $D$ into a union of small cubes $\Da_p$
centered at points $y^{(p)}$ and having no common interior points. The
side of $\Da_p$ is $b\gg a$. The number $\EN(\Da_p)$ of small particles
in $\Da_p$ by formula \eq{25} is:
\begin{equation}
\label{E49}
\EN(\Da_p) = \frac{1}{a^{3\ka_1}} \int_{\Da _p}N(x)dx [1 + o(1)] =
\frac{1}{a^{3\ka_1}}\, N(y^{(p)})\, \n \Da_p\n\, [1 + o(1)],
\end{equation}
where $o(1)$ in the second equation tends to zero as ${\rm diam} \,
\Da_p$ tends to zero, and $\n \Da_p\n$ is the volume of the cube $\Da_p$.
Write the sum in equation \eq{48} as
\begin{eqnarray}
\nn \lefteqn{ \sum_p G(x,y^{(p)}) \, 4\pi h(y^{(p)})
u_e(y^{(p)})a^{2-\ka}\sum_{x_m\in \Da_p}1 } \\
&& = \sum_p G(x,y^{(p)})\, 4\pi h(y^{(p)})\, u_e(y^{(p)})\,
\frac{a^{2-\ka}}{a^{3\ka_1}}\, N(y^{(p)})\n\Da_p\n\f 1 + o(1)\g.\label{E50}
\end{eqnarray}
The sum in \eq{50} is a Riemannian sum for the integral
\begin{equation}
\label{E51}
\int_D G(x,y)\, 4\pi h(y)\, N(y)\, u_e(y)dy.
\end{equation}
The limit of the sum in  \eq{50},  as $a\to 0$, exists if and only if 
$2-\ka=3\ka_1 $, i.e., $\ka_1=(2-\ka)/3$. Note that if $0<\ka<1$, then 
$0<\ka_1<2/3$.

In the region $\n x-x_m\n \geq d$ one has:
\begin{equation}
\label{E52}
\n u_M(x) -u_e(x)\n \leq O\Ba \frac{a}{a^{\ka_1}}\Bz = o(1),\quad a \ra 0.
\end{equation}
The number of small particles in a unit cube is $O\f \frac{1}{d^3}\g =
O\f \frac{1}{a^{3\ka_1}}\g$ if $d = O(a^{\ka_1})$, where $d$ is the
distance between two neighboring particles.
\par
We assume that the functions $h(y), u_e(y)$ and $G(x,y)$ are continuous
functions of $y$, so the error of replacing, for example, $h(y_m)$ by
$h(y^{(p)})$, where $y_m\in \Da_p$, goes to zero as ${\rm diam}\, \Da_p
\ra 0$. The function $G(x,y)$ is not continuous as $y\ra x$, but
$G(x,y)$ is absolutely integrable, so one may remove a small
neighborhood of the singular point $x$ in the integral \eq{51} and the
change of this integral will be negligible if the neighborhood is
sufficiently small. The function $h$ is at our disposal, and we choose
it to be continuous. 
The continuity of $u_e$ and of its limit $u$ follows from
the relation $u\in H^2_{loc}(\R^3)$. A more detailed argument
is given in \cite{R529}.

Assuming that $\ka_1=(2-\ka)/3$ and
passing to the limit $a\ra 0$ in \eq{48}
yields equation \eq{23}. Applying to this equation the operator $L_0$
and using equations \eq{2} and \eq{10}, one gets equation \eq{18} with
$q(x)$ given by \eq{20} and $p(x)$ given by \eq{24}.
\par
We have proved all the claims in the answer to question 1).\qed
$$ $$
Note that if $\ka <0$, then the impedance parameter $ha^{-\ka}$ tends to 
zero as $a\to 0$. 
\ek

Case 2). \ Let us justify the answer to question 2). We assume now that 
$\ka > 
1$. Then \eq{39} implies
\begin{equation}
\label{E53}
Q_m = - 4\pi u_e(x_m)a [1 + o(1)],\quad a \ra 0,
\end{equation}
and equation \eq{43} yields
\begin{equation}
\label{E54}
ca^\gm = - \frac{u_e(x_m)}{a}\,,
\end{equation}
so 
$$\gm = -1, \quad c = -u_e(x_m),$$ 
and
\begin{equation}
\label{E55}
\si_m = - \frac{u_e(x_m)}{a}\, [(1 + o(1)], \quad a \ra 0.
\end{equation}
Let us check when the relation \eq{34} holds, i.e., when formula \eq{35}
is valid, in other words, when formula \eq{35} yields an accurate
approximation of $u_M$, defined by formula \eq{31}. From 
\eq{33} and \eq{53} we conclude, using the relation $d = O(a^{\ka_1})$, 
that
\begin{equation}
\label{E56}
\n G(x,x_m)Q_m\n = O\f \frac{a}{d}\g = O(a^{1-\ka_1}),\quad a \ra 0,\quad
\n x-x_m\n \geq d.
\end{equation}
Furthermore, using \eq{55} and \eq{33}, one
gets:
\begin{equation}
\label{E57}
J \leq O\f \frac{a}{a^{2\ka_1}}\, a^{2-1}\g = O(a^{2-2\ka_1}),\quad a
\ra 0.
\end{equation}
The relation \eq{34} holds if
$ a^{2-2\ka_1}\ll a^{1-\ka_1}$, that is, if $\ka_1 < 1$.
\par
Let us assume that $\ka_1 < 1$, so that formula \eq{35} is applicable. 
We repeat the arguments given below formula \eq{48}. Due to 
formula \eq{53}, now formula \eq{48} takes the form:
\begin{equation}
\label{E58}
u_M(x) = u_0(x) - \sum^M_{m=1}G(x,x_m)\, 4\pi u_e(x_m)a.
\end{equation}
We conclude from this formula that $u_e(x)$ tends to the limit $u(x)$, and
$u$ solves the equation:
\begin{equation}
\label{E59}
u(x) = u_0(x) - \int_D G(x,y)\, 4\pi N(y)\, u(y)dy,
\end{equation}
provided that
\begin{equation}
\label{E60}
\ka_1 = \frac{1}{3}\,,
\end{equation}
and $N(x)$ is defined by the formula \eq{25} for any subdomain $\Da
\subset D$, where $\EN(\Da)$ is the number of small particles in $\Da$.

Applying the operator $L_0$ to \eq{59} one gets equation \eq{18} for
$u(x)$, with $q(x)$ given by \eq{20}, $q_0(x)$ given by \eq{21}, and
$p(x)$ given by the formula
\begin{equation}
\label{E61}
p(x) = 4\pi N(x).
\end{equation}
Since $N(x)\geq 0$, the function $p(x)$ is nonnegative. 
\par
The assumption $\ka > 1$ leads to the equation \eq{18} with
the potential $q(x)$ which can vary much less than in the case $\ka \leq 
1$, because
the function $h(x)$  does not enter in the
definition of $q(x)$ when $\ka > 1$.
\section{Creating materials with a desired refraction coefficient}
If $\ka < 1$ and $\ka_1 = (2-\ka)/3\,$, then equations \eq{18},
\eq{20} and \eq{24} hold. Thus, given $n_0^2(x)$ and $n^2(x)$, one
calculates
\begin{equation}
\label{E62}
p(x) = k^2[n_0^2(x)-n^2(x)] := p_1(x) + ip_2(x).
\end{equation}
From \eq{24} and \eq{62} one gets an equation for finding $h(x):= h_1(x)
+ ih_2(x)$ and $N(x)\geq 0$:
\begin{equation}
\label{E63}
4\pi [h_1(x) + ih_2(x)]N(x)= p_1(x) + ip_2(x).
\end{equation}
Thus 
\begin{equation}
\label{E64}
N(x)h_1(x) = \frac{p_1(x)}{4\pi}\,,\quad N(x)h_2(x) = \frac{p_2(x)}{4\pi}\,.
\end{equation}
There are many solutions $\{ h_1,h_2,N\}$ of two equations \eq{64} for
the three unknown functions $h_1,h_2,N(x)$, $h_2\leq 0$, $N\geq 0$. The 
condition $\Im n^2(x)
\geq 0$ implies $\Im p = p_2 \leq 0$, which agrees with th inequalities
$h_2\leq 0$, $N\geq 0$. 
%Clearly, $N(x)\geq 0$. 
One takes $N(x) = h_1(x) = h_2(x) = 0$ at the points at
which $p_1(x) = p_2(x) = 0$. At the points at which $\n p(x)\n > 0$, one
may take 
\begin{equation}
\label{E65}
N(x) = N = {\rm const}\,,\quad h_1(x) = \frac{p_1(x)}{4\pi N}\,,\quad
h_2(x) = \frac{p_2(x)}{4\pi N}\,.
\end{equation}
Let us partition $D$ into a union of small cubes $\Da_p$, which have no
common interior points, and which are centered at the points $y^{(p)}$, 
and embed
in each cube $\Da_p$ the number
\begin{equation}
\label{E66}
\EN(\Da_p) = \Bka \frac{1}{a^{2-\ka}}\int_{\Da_p}N(x)dx \Bkz
\end{equation}
of small balls $D_m$ of radius $a$, centered at the points $x_m$, where
$[c]$ stands for the integer nearest to $c > 0$. Let us put these balls 
at
the distances $O\f a^{\frac{2-\ka}{3} }\g$, and prepare the boundary
impedances of these balls equal to $\frac{h(x_m)}{a^\ka}\,$.

{\it  Then the
resulting material, which is obtained by embedding small particles into 
$D$ by the above recipe,  will have the desired refraction coefficient 
$n^2(x)$ with an error going to zero as $a\ra 0$.}

\section{Conclusions}\label{S4}
Wave scattering by many small particles, embedded into a material with a 
known refraction coefficient, is studied in this Letter under various 
assumptions about the orders $\ka_1$ and $-\ka$
 with respect to powers of $a$ of the
distances $d=O(a^{\ka_1})$ between the neighboring small particles and 
their boundary impedances $\zeta=O(a^{-\ka})$. 
\par
If $a$ is the characteristic size of a spherical small particle $D_m$, 
$d=O(a^{\ka_1})$ is the distance between the neighboring particles, 
$\zeta_m=\frac{h(x_m)}{a^\ka}$ is the boundary impedance, and $x_m$ is 
the center of $D_m$, then the equations are derived, as $a \to 0$, for the 
effective field 
in the medium, consisting of many small particles, 
embedded in a given material (according to the recipe, derived in Section 
2),  under the following assumptions:
\begin{zael}
\item $0< \ka \le 1,$ $\ka_1=\frac{2-\ka}{3}$, and \eq{25} holds,
\item $\ka > 1$, $\ka_1=\frac{1}{3}$, and \eq{25} holds.
\end{zael}
A remark is made about the case $\ka_1=1$ and $\ka=-1$. In this case 
$d=O(a)$ is of the order of the size of a small particle.

The results ar used for formulating a recipe for creating materials with a 
desired refraction coefficient. This recipe is similar to the
one, given in \cite{R527},
\cite{R528} in the case $\ka_1=1/3$ and $\ka=1$.
\end{document}